\documentstyle[11pt,newpasp,twoside]{article}
\begin{document}
\title{Numerical hydrodynamics: SPH vs. AMR}
\author{Tomek Plewa}
\affil{Nicolaus Copernicus Astronomical Center, Bartycka 18, 00-716 Warsaw,
Poland}
\begin{abstract}

The advantages and disadvantages of two approaches to astrophysical
hydrodynamics, Smoothed Particle Hydrodynamics and Adaptive Mesh
Refinement, are briefly discussed together with some current
problems of computational hydrodynamics.

\end{abstract}
\section{From Henyey to SPH and AMR}
A history of numerical hydrodynamics in application to a problem of
protostellar collapse begins more than three decades ago. It is
remarkable to note that already at that time numerical simulations
followed two distinct paths. Bodenheimer \& Sweigart (1968) and
Bodenheimer (1968) used a Lagrangian implicit method closely following
Henyey's approach (Henyey, Forbes \& Gould 1964) commonly used for
stellar evolution computations. On the other hand, Larson (1969)
developed a variant of an Eulerian scheme for radiation hydrodynamics
in the optically thick regime. Despite the use of very different
numerical schemes, both studies yielded much similar results.

However, problems like the massive removal of angular momentum from
the clump of molecular gas collapsing into stellar densities, could
not be addressed within the one-dimensional framework. Among the
simplest ways of removing excess angular momentum is the deposition of
angular momentum in the orbital motion of planets formed around a
central star, or by the formation of a multiple stellar
system. Neither of the two ways of removal of angular momentum can be
studied in detail in one dimension. To overcome this problem Lucy
(1977) proposed a method in which the distribution of hydrodynamical
quantities is approximated with a discrete set of diffuse clouds of
matter, Smoothed Particle Hydrodynamics (Monaghan 1992). The SPH
method is purely Lagrangian, can be easily implemented on a computer,
and allows for performing 3-dimensional simulations even on relatively
small systems.

On the other hand, the solution of the hydrodynamical equations in
multidimensions poses severe problems for grid based codes. In the
Lagrangian approach the grid cells follow individual fluid elements
and whenever a significant amount of shear is present in the flow the
numerical grid becomes heavily distorted. This problem is nicely
illustrated by the work of Woodward (1976) in his study of a shock-cloud
interaction problem. Since a strong distortion of the computational grid
leads to a loss of accuracy of the discretization, most of the later
research on Lagrangian schemes was devoted to the development of
sophisticated rezoning modules which would allow to adequately follow
individual fluid elements with an accuracy independent of their shapes.

An overall quality of the solution in the Eulerian model is determined by
both the order of the advection scheme and the resolution provided by
the numerical grid. Woodward and Colella (1984) demonstrated that
substantial savings in terms of both memory and CPU time can be
achieved when advection schemes of high-order are used instead of
their low-order counterparts. These modern advection algorithms,
shock-capturing schemes, gained much popularity during the last decade
and follow the seminal works of Godunov (1959), van Leer (1976) and Harten
(1983). The shock-capturing methods are characterized by at least
a second order accuracy in the smooth part of the flow, with flow
discontinuities being resolved in one or two zones, and no artificial
viscosity is required to obtain a physically correct solution.

It took almost 20 years for Woodward's pioneering study of a
shock-cloud interaction to be successfully addressed with an Eulerian
code (Klein, McKee \& Colella 1994). However, even the use of a
shock-capturing scheme was still not enough to adequately resolve the
flow structure and to obtain a convergent solution: a novel technique
of local adaptive mesh refinement (AMR; Berger \& Colella 1989) was
employed. In essence, AMR is a way of efficient discretization which
allows to concentrate the computational effort in regions in which the
errors of the solution are large. From a conceptual point of view AMR
is similar to multigrid techniques with local refinements (Brandt 1984),
in which the computational volume is covered with a hierarchy of
completely nested grid patches and the resolution increases between
parent and child grid levels by some predefined integer
factor. However, due to the conservative character of the time-dependent
hydrodynamical equations, the AMR method has to fulfill additional
requirements making its actual computer implementation, to say the
least, difficult.
\section{Far side of numerical modelling}
The planning of numerical simulations is much similar to the concept
of project design known from industry rather than science. However, in
both cases, given that objectives of the project have been clearly
defined, a proper choice of the method of solution is necessary.  Once
an adequate tool is available one needs to find a way for using it. In
practice this demand translates into having access to a computer
installation which has to provide enough resources for obtaining
the solution to the problem at hand. At first this last issue may appear
to be a far side of numerical modelling, but to those involved in
computationally intensive calculations this is that side which
ultimately makes numerical simulation possible to accomplish.

Large simulations represent a separate class of computational problems
which are characterized by a high value of the product of computer
memory and processor time. This practical definition follows the
observation that simulations done on grids with just a few thousand
zones (or particles) might be demanding once the modelled system has
to be evolved for a long time. A typical example here are simulations
of accretion disks which suffer from short evolutionary timescales
present near the inner edge of the disk. On the other hand, large
memory is required for virtually all three-dimensional problems, even
with memory savings indirectly offered by modern advection methods, or
whenever very high, even if only locally, resolution is needed. The
case of planet formation in protoplanetary disks serves here as an
example.

Furthermore, the large size of the simulation demands a certain
minimum speed of the processor and the main memory: program data has
to be accessed and subsequently processed within acceptable time. Most
of the large simulations put much more strain on the memory subsystem,
including the system bus, than on the processor and, therefore, are
``bandwidth'' rather than CPU limited. Certainly, not only some of
those available but even the largest existing computer installation
may not satisfy the requirement imposed by a ``memory times processing
speed'' limit: numerical experiments are limited by the availability
of computer resources.  Finally, also storage and analysis of large
($\sim 10^2$ times problem memory) data sets requires specialized
hardware and software.

The variety of existing computer architectures does not necessarily
make numerical simulations easier to plan and conduct. The performance
of a computer code may differ vastly between scalar and vector
architectures. Distributed and shared memory machines follow different
programming paradigms, may ask for deep modifications to the existing
codes. Such ``technicalities'' should be considered at the very
beginning of any project which involves intensive computations as they
might be difficult or impossible to overcome later.  Keeping an eye
wide open on current trends in computer architecture design, usually
reflected in the policy of supercomputer centers, seems to be nowadays
as important for numerical modelers as the development and application
of new algorithms.
\section{A few words of critics}
In my introductory remarks I sketched a path that astrophysical
hydrodynamics followed from relatively simple one-dimensional
Lagrangian models to full three-dimensional simulations done with help
of the SPH and AMR methods. I find this development remarkable even
though, as we could learn during this conference, after three decades
the problem of one-dimensional stellar collapse may still offer enough
of material for debates. As for the comparison between SPH and AMR,
the two so extremely distinct approaches to hydrodynamics, there appears
to exist a sharp boundary dividing the groups of users of these two
methods. However, both methods have their own limits and aspects which
ask for special attention. I will list them briefly.

The Lagrangian nature of the SPH method makes spatial resolution
impossible to maintain comparable in all regions of the computational
volume, a worry for those who aim at studying properties of voids in
cosmological simulations. Resolution near the flow discontinuities is
usually poor, equivalent to a few smoothing lengths ($\sim$ particle
``radii''), in most implementations of the SPH method shocks are
handled with help of an artificial viscosity. Also the resolution and
stability (in multidimensions) of contact discontinuities is
problematic. SPH particles may ``penetrate'' each other increasing the
diffusivity of the scheme and making studies which involve mixing and
shear flows difficult.

Contrary to SPH, the resolution of an Eulerian model is defined by the
smallest spatial scale and, would AMR not have been invented, no
Eulerian calculation could achieve a resolution in mass comparable to
that of large SPH simulations. However, the nonuniformity of the
spatial resolution makes the AMR method unsuitable for studying
turbulent flows which demand a smooth sampling of the spatial
scales. Also, the roughness of the AMR grids adds a certain amount of
vorticity to the model since each boundary between the levels in the
AMR hierarchy appears as an obstacle to the flow and acts as a source
of purely spurious vorticity.

Last but not least, the numerically ``correct'' results are those
which correspond to the convergent solution. The Eulerian routine of
doubling the resolution from one model to another and comparing low-
and high-resolution models does not translate into doubling the number
of particles in a SPH simulation.  A concept of the smoothing length,
usually mentioned when the spatial resolution of the SPH model is
presented, is in this respect misleading since a twofold decrease of
the smoothing length requires an increase in the number of
neighbouring particles by a factor of $2^D$, where $D$ is the number
of spatial dimensions. Yet, in most cases convergence studies (if
present at all!) are done with the number of particles being varied by
a factor of just a few.

I decided not to mention problems related to radiation transport or
the incorporation of magnetic fields into hydrodynamical simulations.
I did it on purpose since their complexity will likely add to the
confusion; it is better to leave out their discussion for the years to
come and closely follow the development of more efficient and correct
algorithms. At this very moment it is my conviction that even pure
hydrodynamics still poses challenging problems and obtaining
physically meaningful results is, and as it has ever been, a difficult
task.
\acknowledgements{I thank the organizing committee for financial
support and hospitality during the conference. This work was partly
supported by grant 2.P03D.014.19 from the Polish Committee for
Scientific Research.}
\end{document}